\documentclass[12pt]{article}
\usepackage[dvips]{graphicx}

\begin{document}

\title{Guest Charge and Potential Fluctuations in Two-Dimensional
Classical Coulomb Systems}

\author{Bernard Jancovici$^1$ and Ladislav {\v S}amaj$^{1,2}$}

\maketitle

\begin{abstract}
A known generalization of the Stillinger-Lovett sum rule for a guest
charge immersed in a two-dimensional one-component plasma (the second moment
of the screening cloud around this guest charge) is more simply
retrieved, just by using the BGY hierarchy for a mixture of several species;
the zeroth moment of the excess density around a guest charge immersed in a
two-component plasma is also obtained. The moments of the electric potential
are related to the excess chemical potential of a guest charge; explicit
results are obtained in several special cases.
\end{abstract}

\medskip


\vfill

\noindent $^1$
Laboratoire de Physique Th\'eorique, Universit\'e de
Paris-Sud, B\^atiment 210, 91405 Orsay Cedex, France
(Unit\'e Mixte de Recherche no. 8627 - CNRS); 
E-mail: Bernard.Jancovici@th.u-psud.fr

\noindent $^2$ 
Institute of Physics, Slovak Academy of Sciences, D\'ubravsk\'a cesta 9, 
\newline 845 11 Bratislava, Slovak Republic; 
E-mail: Ladislav.Samaj@savba.sk

\newpage

\renewcommand{\theequation}{1.\arabic{equation}}
\setcounter{equation}{0}

\section{Introduction}
One of us (L. \v{S}.) has derived a generalization of the Stillinger-Lovett sum
rule for a guest charge immersed in a two-dimensional one-component plasma
\cite{Samaj07}: an exact simple expression for the second moment of the 
screening cloud around the guest charge was obtained, 
by using a mapping technique onto a discrete one-dimensional 
anticommuting-field theory. In the present paper, we first show that the same
result can be obtained  in a simpler way by just using the BGY hierarchy,
which provides also more general results. 

The excess chemical potential of a guest charge (which can be
expressed in terms of the charge density of the screening cloud) has an
expansion in powers of the guest-particle charge $Ze$, which  allows to
compute the average of powers (moments) of the electric potential.

We consider a classical (i.e. non-quantum) system of charged particles
located in an infinite two-dimensional (2D) plane of points 
${\bf r}\in {\rm R}^2$. According to the laws of 2D electrostatics, the
particles can be thought of as infinitely long charged lines in the 3D which
are perpendicular to the 2D plane.
The electrostatic potential $v$ at a point ${\bf r}$, induced by
a unit charge at the origin ${\bf 0}$, is thus given by the 2D
Poisson equation
\begin{equation} \label{1.1}
\Delta v({\bf r}) = - 2\pi \delta({\bf r}) .
\end{equation}
The solution of this equation, subject to the boundary condition
$\nabla v({\bf r})\to 0$ as $\vert {\bf r}\vert \to \infty$, reads
\begin{equation} \label{1.2}
v(r) = - \ln\left( \frac{r}{L} \right) ,
\end{equation}
where $r=\vert {\bf r}\vert$ and the free length constant $L$,
which determines the zero point of the potential, will be set
for simplicity to unity.
The Fourier component of this potential $\tilde{v}({\bf k})\propto 1/k^2$
exhibits the characteristic singularity at $k=0$, which maintains
many generic properties (like screening) of ``real'' 3D charged systems.

A general Coulomb system consists of $M$ mobile species
$\alpha = 1,2,\ldots,M$ with the corresponding charges $e_{\alpha}$
(which may be integer multiples of the elementary charge $e$). Mobile
particles may be embedded in a fixed uniform background 
of charge density $\rho_b$. The most studied models are the one-component
plasma (OCP), which corresponds to $M=1$ with $e_1=e$ and $\rho_b$ of opposite
sign, and the symmetric two-component plasma (TCP), which
corresponds to $M=2$ with $e_1=e$, $e_2=-e$, $\rho_b=0$. 
The interaction energy of a configuration $\{ {\bf r}_i,e_{\alpha_i}\}$
of the charged particles plus the background is
\begin{equation} \label{1.3}
E = \sum_{i<j} e_{\alpha_i} e_{\alpha_j} v(\vert {\bf r}_i-{\bf r}_j\vert)
+ \sum_i e_{\alpha_i} \phi_b({\bf r}_i) + E_{b-b} ,
\end{equation}
where $\phi_b({\bf r})$ is the one-body potential created by
the background and the background-background energy term $E_{b-b}$
does not depend on the particle coordinates.
In the case of point particles, for many-component systems with at least
two oppositely species, the singularity of the Coulomb potential
(\ref{1.2}) at the origin ${\bf r}={\bf 0}$ prevents, for small enough
temperatures, the thermodynamic stability against the collapse of
positive-negative pairs of charges.
In those cases, one introduces to $v$ a short-range repulsion
which prevents the collapse.

The Coulomb system is considered in thermodynamic equilibrium,
at inverse temperature $\beta=1/(k_{\rm B}T)$.
The thermal average over an infinite neutral system will be denoted by
$\langle \cdots \rangle$.
In terms of the microscopic density of particles of species $\alpha$,
$\hat{n}_{\alpha}({\bf r}) = \sum_i \delta_{\alpha,\alpha_i}
\delta({\bf r}-{\bf r}_i)$, the microscopic total number density and
the microscopic total charge density are defined, respectively, by
\begin{equation} \label{1.4}
\hat{n}({\bf r}) = \sum_{\alpha} \hat{n}_{\alpha}({\bf r}) , \qquad
\hat{\rho}({\bf r}) = \sum_{\alpha} e_{\alpha} \hat{n}_{\alpha}({\bf r})
+\rho_b.
\end{equation}
The microscopic electrostatic potential created by the particle-background
system at point ${\bf r}$ is given by
\begin{equation} \label{1.5}
\hat{\phi}({\bf r}) = \int {\rm d}{\bf r}' v({\bf r}-{\bf r}')
\hat{\rho}({\bf r}') .
\end{equation}
At the one-particle level, the homogeneous number density of species $\alpha$
and the total particle number density are given respectively by
\begin{equation} \label{1.6}
n_{\alpha} = \langle \hat{n}_{\alpha}({\bf r}) \rangle , \qquad
n = \langle \hat{n}({\bf r}) \rangle .
\end{equation} 
The charge density $\rho=\langle \hat{\rho}({\bf r})\rangle$ vanishes
due to the charge neutrality of the system.
At the two-particle level, one introduces the translationally invariant
two-body densities
\begin{eqnarray}
n_{\alpha\alpha'}^{(2)}(\vert {\bf r}-{\bf r}'\vert) & = &
\left\langle \sum_{i\ne j} \delta_{\alpha,\alpha_i} \delta({\bf r}-{\bf r}_i)
\delta_{\alpha',\alpha_j} \delta({\bf r}'-{\bf r}_j) \right\rangle
\nonumber \\ & = &
\langle \hat{n}_{\alpha}({\bf r}) \hat{n}_{\alpha'}({\bf r}') \rangle
- \langle \hat{n}_{\alpha}({\bf r}) \rangle \delta_{\alpha,\alpha'}
\delta({\bf r}-{\bf r}') . \label{1.7}
\end{eqnarray}
It is useful to consider also the pair distribution functions
\begin{equation} \label{1.8}
g_{\alpha\alpha'}(\vert {\bf r}-{\bf r}'\vert) =
\frac{n^{(2)}_{\alpha\alpha'}(\vert {\bf r}-{\bf r}'\vert)}{
n_{\alpha} n_{\alpha'}},
\end{equation}
the (truncated) pair correlation functions
$h_{\alpha\alpha'}=g_{\alpha\alpha'}-1$, as well as the three-body analogous
quantities
\begin{equation} \label{1.9}
g_{\alpha\alpha'\alpha''}^{(3)}({\bf r},{\bf r}',{\bf r}'') =
\frac{n^{(3)}_{\alpha\alpha'\alpha''}({\bf r},{\bf r}',{\bf r}'')}
{n_{\alpha} n_{\alpha'} n_{\alpha''}}
\end{equation}
and the (truncated) three-body correlation function
\begin{eqnarray}
h_{\alpha\alpha'\alpha''}^{(3)}({\bf r},{\bf r}',{\bf r}'') & = &
g_{\alpha\alpha'\alpha''}^{(3)}({\bf r},{\bf r}',{\bf r}'')
-h_{\alpha\alpha'}(\vert {\bf r}-{\bf r}'\vert) 
-h_{\alpha'\alpha''}(\vert {\bf r'}-{\bf r}''\vert) \nonumber \\ & &
-h_{\alpha''\alpha}(\vert {\bf r''}-{\bf r}\vert)-1. \label{1.10}
\end{eqnarray}

The paper is organized as follows. In Section 2, we use the BGY hierarchy for
studying the general mixture of $M$ species of mobile particles embedded in a
fixed uniform background. By taking the limit of one of the densities going
to zero, we get the case of a guest charge. We retrieve the second moment of
the screening cloud around a guest charge immersed in an OCP; we get also the
zeroth moment of the excess total number density around a guest charge
immersed in a TCP. In Section 3, the general formalism for relating the
moments of the electric potential to the excess chemical potential of a guest
charge is established. The following Sections study special cases when
explicit calculations are possible: the high-temperature (Debye-H\"{u}ckel)
limit in Section 4, the OCP at $\beta e^2=2$ in Section 5, the TCP in 
Section 6. Section 7 is a Conclusion.

\renewcommand{\theequation}{2.\arabic{equation}}
\setcounter{equation}{0}

\section{Sum rules for a guest charge immersed in a
Coulomb system}
We wish to rederive and extend the result of \cite{Samaj07} about the 2D OCP in
which a point guest charge $Ze$ is immersed. Let the charge density at
${\bf r}$ knowing that there is a guest charge $Ze$ at the origin be
$\rho({\bf r}\vert Ze,{\bf 0})$. In \cite{Samaj07}, 
its second moment was shown to be
\begin{equation} \label{2.1}
\int {\rm d}{\bf r}\:r^2 \rho({\bf r}\vert Ze,{\bf 0}) =
-\frac{2}{\pi\beta e n}\left[Z\left(1-\frac{\beta e^2}{4}\right)
+Z^2\frac{\beta e^2}{4}\right]. 
\end{equation}
Our rederivation uses only the BGY hierarchy.

\subsection{General sum rule for a mixture with a background} 
We start with the mixture of $M$ mobile species, with a fixed uniform
background, described in the Introduction. Finally, we shall
consider a mixture of only 2 species with respective charges $e_1=e$ and
$e_2=Ze$; at the end, the density $n_2$ will be chosen as 0, leaving only one
guest charge. But for being able to consider the TCP as well, we start with
the more general case of $M$ mobile species. The neutrality constraint is
\begin{equation} \label{2.2}
\sum_{\alpha}n_{\alpha}e_{\alpha} = -\rho_b . 
\end{equation}

The mixture with a background has been studied in three dimensions 
by Suttorp and van Wonderen~\cite{Suttorp}. 
They used the BGY hierarchy and thermodynamical properties of the system 
for deriving, among other things, a second-moment sum rule, 
which however involves some thermodynamical functions (the partial
derivatives of each density $n_{\alpha}$ with respect to the background
density $n_b$); there is no explicit expression for these partial derivatives.
Fortunately, we found that, in two dimensions, the formalism becomes much
simpler and only the BGY hierarchy has to be used (the thermodynamical
properties are no longer involved).

The second equation of the BGY hierarchy \cite{Martin}, with 
$h_{\alpha\alpha'}(r)$ the correlation function between a particle of
species $\alpha$ at ${\bf r}$ and a particle of species $\alpha'$ at the
origin, is
\begin{eqnarray}
\beta^{-1}\nabla h_{\alpha\alpha'}(r) & = & 
\nonumber \\
&-&\sum_{\alpha''}n_{\alpha''}\int {\rm d}{\bf r}''\:h_{\alpha'\alpha''}
(r'')e_{\alpha}e_{\alpha''}
\nabla v(\vert {\bf r}-{\bf r}''\vert) \nonumber \\ 
&-&h_{\alpha\alpha'}(r)
e_{\alpha}e_{\alpha'}\nabla v(r)
-e_{\alpha}e_{\alpha'} \nabla v(r) \nonumber \\
&-&\sum_{\alpha''}n_{\alpha''}\int {\rm d}{\bf r}''\:
h_{\alpha\alpha'\alpha''}^{(3)}
({\bf r},{\bf 0},{\bf r}'')e_{\alpha}e_{\alpha''}
\nabla v(\vert {\bf r}-{\bf r}''\vert). \label{2.3} 
\end{eqnarray}
The integral in the first term in the rhs of (\ref{2.3}) is proportional to
the electric field at ${\bf r}$ due to the charge distribution
$h_{\alpha'\alpha''}$ which has a circular symmetry around the origin. 
Thus, using Newton's theorem, one can rewrite this integral as
\begin{equation} \label{2.4}
\int {\rm d}{\bf r}''\:h_{\alpha'\alpha''}(r'')
\nabla v(\vert {\bf r}-{\bf r}''\vert)=
\nabla v(r)\int_{r''<r}{\rm d}{\bf r}''\:h_{\alpha'\alpha''}(r''). 
\end{equation}
The integral in the rhs of (\ref{2.4}) can be written as
$\int_{r''<r}\ldots =\int\ldots-\int_{r''>r}\ldots$ and the perfect screening 
of the charge $e_{\alpha'}$ gives \cite{Gruber}
\begin{equation} \label{2.5}
\sum_{\alpha''}e_{\alpha''}n_{\alpha''}\int {\rm d}{\bf r}''\:
h_{\alpha'\alpha''}(r'')=-e_{\alpha'}.
\end{equation}
Therefore (\ref{2.3}) can be rewritten as
\begin{eqnarray}
\beta^{-1}\nabla h_{\alpha\alpha'}(r) & = &
\nonumber \\
&e_{\alpha}&\sum_{\alpha''}n_{\alpha''}e_{\alpha''}\nabla v(r) 
\int_{r''>r}{\rm d}{\bf r}''\:h_{\alpha'\alpha''}(r'') \nonumber \\
&-&h_{\alpha\alpha'}(r)e_{\alpha}e_{\alpha'}\nabla v(r) \nonumber \\
&-&e_{\alpha}\sum_{\alpha''}n_{\alpha''}\int {\rm d}{\bf r}''\:
h_{\alpha\alpha'\alpha''}^{(3)}({\bf r},{\bf 0},{\bf r}'')e_{\alpha''}
\nabla v(\vert {\bf r}-{\bf r}''\vert). \label{2.6}
\end{eqnarray}

In order to make a second moment to appear, we take the scalar product of both
sides of (\ref{2.6}) with ${\bf r}$ and integrate on 
${\bf r}$. Integrating by parts the lhs and performing the
integration on ${\bf r}$ first in the first term of the rhs, one
finds 
\pagebreak 
\begin{eqnarray}
&&-2\beta^{-1}\int {\rm d}{\bf r}\:h_{\alpha\alpha'}(r)=
-\pi e_{\alpha}\sum_{\alpha''}n_{\alpha''}e_{\alpha''}\int{\rm d}{\bf r}\:r^2
h_{\alpha'\alpha''}(r) \nonumber \\
& &+e_{\alpha}e_{\alpha'}\int{\rm d}{\bf r}\:h_{\alpha\alpha'}(r) 
\label{2.7} \\
& &+e_{\alpha}\sum_{\alpha''}n_{\alpha''}e_{\alpha''}
\int{\rm d}{\bf r}\:{\rm d}({\bf r}''-{\bf r})\:
h_{\alpha\alpha'\alpha''}^{(3)}({\bf r},{\bf 0},{\bf r}'')
\frac{({\bf r}-{\bf r}'')\cdot{\bf r}}{({\bf r}-{\bf r}'')^2}
\nonumber
\end{eqnarray}
(in the last term, since $h^{(3)}$ depends only on ${\bf r}$ and the 
difference ${\bf r}''-{\bf r}$, we have replaced the integration
on ${\bf r}''$ by an integration on ${\bf r}''-{\bf r}$).
An important simplification has occurred in 2D where ${\bf r}\cdot\nabla v(r)$
has the constant value $-1$, while in three dimensions, with the potential
$v(r)=1/r$, one finds $-v(r)$, a result which has led to a more complicated
calculation in \cite{Suttorp}.

Now, we multiply both sides of (\ref{2.7}) by $n_{\alpha}$ and sum on $\alpha$.
The term involving $h^{(3)}$ can be simplified by using symmetries under 
permutations of the variables. Indeed, $h^{(3)}$ has the symmetry property
\begin{equation}
h_{\alpha\alpha'\alpha''}^{(3)}({\bf r},{\bf 0},{\bf r}'')= 
h_{\alpha''\alpha'\alpha}^{(3)}({\bf r}'',{\bf 0},{\bf r}). \label{2.8}
\end{equation} 
Thus, interchanging the summation variables $\alpha$ and $\alpha''$, and the
variables ${\bf r}$ and ${\bf r}''$, we obtain
\begin{eqnarray}
& &\sum_{\alpha,\alpha''}n_{\alpha}e_{\alpha}n_{\alpha''}e_{\alpha''}
\int{\rm d}({\bf r}''-{\bf r})\:h_{\alpha\alpha'\alpha''}^{(3)}
({\bf r},{\bf 0},{\bf r}'')
\frac{({\bf r}-{\bf r}'')\cdot{\bf r}}{({\bf r}-{\bf r}'')^2}
\nonumber \\
& &=\sum_{\alpha,\alpha''}n_{\alpha}e_{\alpha}n_{\alpha''}e_{\alpha''}
\int{\rm d}({\bf r}''-{\bf r})\:h_{\alpha\alpha'\alpha''}^{(3)}
({\bf r},{\bf 0},{\bf r}'') 
\frac{({\bf r}''-{\bf r})\cdot{\bf r}''}{({\bf r}-{\bf r}'')^2}
\nonumber \\
&&=\frac{1}{2}\sum_{\alpha,\alpha''}n_{\alpha}e_{\alpha}
n_{\alpha''}e_{\alpha''}
\int{\rm d}({\bf r}''-{\bf r})\:h_{\alpha\alpha'\alpha''}^{(3)}
({\bf r},{\bf 0},{\bf r}''),  \label{2.9} 
\end{eqnarray}
where the last line is the half sum of the two first ones. 

Using (\ref{2.9}) in (\ref{2.7}) gives
\begin{eqnarray}
& &-2\beta^{-1}\sum_{\alpha}n_{\alpha} \int{\rm d}{\bf r}\:h_{\alpha\alpha'}(r)
=\pi\rho_b\sum_{\alpha''} n_{\alpha''}e_{\alpha''}\int{\rm d}{\bf r}\:r^2
h_{\alpha'\alpha''}(r) \nonumber \\
& &+e_{\alpha'}\sum_{\alpha}n_{\alpha}e_{\alpha} \int{\rm d}{\bf r}\:
h_{\alpha\alpha'}(r) \nonumber \\
& &+\frac{1}{2}\sum_{\alpha,\alpha''}n_{\alpha}e_{\alpha}
n_{\alpha''}e_{\alpha''}
\int{\rm d}{\bf r}\:{\rm d}({\bf r}''-{\bf r})\:
h_{\alpha\alpha'\alpha''}^{(3)}({\bf r},{\bf 0},{\bf r}''). \label{2.10}
\end{eqnarray}
For the second term in the rhs of (\ref{2.10}), perfect screening
\cite{Gruber} gives  
$-e_{\alpha'}^2$. 
For the last term in the rhs of (\ref{2.10}), perfect screening gives
\pagebreak
\begin{eqnarray}
& &+\frac{1}{2}\sum_{\alpha,\alpha''}n_{\alpha}e_{\alpha}
n_{\alpha''}e_{\alpha''}
\int{\rm d}{\bf r}\:{\rm d}({\bf r}''-{\bf r})\:
h_{\alpha\alpha'\alpha''}^{(3)}({\bf r},{\bf 0},{\bf r}'') \nonumber \\
& &=-\frac{1}{2}\sum_{\alpha''}n_{\alpha''}e_{\alpha''}
(e_{\alpha'}+e_{\alpha''})\int{\rm d}{\bf r}\:h_{\alpha'\alpha''}(r)
\nonumber \\
& &=\frac{1}{2}\left[e_{\alpha'}^2-\sum_{\alpha''}n_{\alpha''}e_{\alpha''}^2
\int{\rm d}{\bf r}\:h_{\alpha'\alpha''}(r)\right]. \label{2.11}
\end{eqnarray}
Thus (\ref{2.10}) becomes the general second-moment sum rule
\begin{equation} \label{2.12}
-\beta\pi\rho_b\sum_{\alpha}n_{\alpha}e_{\alpha}
\int{\rm d}{\bf r}\:r^2 h_{\alpha'\alpha}(r) 
=\frac{1}{2}\sum_{\alpha}n_{\alpha}(4-\beta e_{\alpha}^2)
\int{\rm d}{\bf r}\:h_{\alpha'\alpha}(r)-\frac{1}{2}\beta e_{\alpha'}^2. 
\end{equation}

By multiplying (\ref{2.12}) by $n_{\alpha'}e_{\alpha'}$ and summing on 
$\alpha'$, one recovers the usual Stillinger-Lovett sum rule \cite{Martin}.
But (\ref{2.12}) is a stronger sum rule.

\subsection{Guest charge in a one-component plasma}
We come to the case of a mixture of two species, with charge $e_1=e$
and density $n_1$, charge $e_2=Ze$ and density $n_2$, respectively. 
We choose $\alpha'=2$ in (\ref{2.12}). 
For dealing with one guest charge $Ze$ only, we set 
$n_2=0$, $n_1=n$, $-\rho_b=ne$; $n$ times the integral of $h_{21}$ is $-Z$, 
by perfect screening. The sum rule (\ref{2.12}) becomes
\begin{equation} \label{2.13}
\beta \pi n^2 e^2\int{\rm d}{\bf r}\:r^2 h_{21}(r)=
-2\left[Z\left(1-\frac{\beta e^2}{4}\right)+Z^2\frac{\beta e^2}{4}\right].
\end{equation}
Since $\rho({\bf r}\vert Ze,{\bf 0})=neh_{21}(r)$, (\ref{2.13}) is (\ref{2.1}).

This result (\ref{2.1}) can also be retrieved by a different method in the
next subsection.

\subsection{Another derivation}
(\ref{2.1}) can be derived in another way if we \emph{assume} that this second
moment can be expanded in integer powers of $Z$.

In the limit of small $Z$, the term linear in Z in (\ref{2.1}) can be obtained
by linear response theory. 
Indeed, if we introduce a guest charge $Ze$, located at the origin,
into an OCP, the additional Hamiltonian is
\begin{equation} \label{A.1}
\hat{H}'=Ze\hat{\phi}({\bf 0}), 
\end{equation}
where $\hat{\phi}({\bf 0})$ is the microscopic electric potential created by
the OCP at the origin. 
To first order in $Z$, the charge density at ${\bf r}$ is 
\begin{equation} \label{A.2}
\rho({\bf r}\vert Ze,{\bf 0})
= -\beta\langle\hat{\rho}({\bf r})Ze\hat{{\phi}}({\bf 0})\rangle^{\rm T}
= -Ze\beta\int{\rm d}{\bf r}'\:v(r')
\langle\hat{\rho}({\bf r})\hat{\rho}({\bf r}')\rangle^{\rm T},  
\end{equation}
where $\langle\cdots\rangle^{\rm T}$ denotes a truncated average. 
We define the Fourier transforms as
\begin{equation} \label{A.3}
\tilde{f}({\bf k}) = \int{\rm d}{\bf r}\:\exp({\rm i}{\bf k}\cdot{\bf r})
f({\bf r}). 
\end{equation}
Then, the Fourier transform of $\rho({\bf r}\vert Ze,{\bf 0})$ is
\begin{equation} \label{A.4}
\tilde{\rho}({\bf k}\vert Ze)=-\beta Ze\frac{2\pi}{k^2}\tilde{S}(k),
\end{equation}
since the Fourier transform of $v(r)$ is $2\pi/k^2$ and the Fourier transform
of the correlation of charge densities is $\tilde{S}(k)$. 
For small $k$, $\tilde{S}(k)$ has the expansion \cite{Martin} 
\begin{equation} \label{A.5}
\tilde{S}(k)=\frac{k^2}{2\pi\beta}-\frac{(1-\beta e^2/4)k^4}{
4\pi^2 n\beta^2 e^2}+\cdots.  
\end{equation}
Therefore, we get the zeroth moment 
\begin{equation} \label{A.6}
\int{\rm d}{\bf r}\:\rho({\bf r}\vert Ze,{\bf 0})=-Ze, 
\end{equation}
in agreement with equation (1.20) in \cite{Samaj07}, 
and the part linear in $Z$ of the second moment (\ref{2.1}).

It may be remarked that the $k^4$ term of (\ref{A.5}) is related to the
compressibility, which is exactly known only for the 2D OCP \cite{Hauge}. 
Therefore, no extension to 3D, with a closed result, seems possible.

In the opposite case of large $Z$, the impurity expels the mobile particles
from a large region around it, leaving only the background. Essentially,
$\rho({\bf r}\vert Ze, {\bf 0})=-ne$ for $r<R$, where $R$ is
some large radius, and $\rho({\bf r}\vert Ze, {\bf 0})=0$ for $r>R$ (there is a
transition region \cite{JancoviciPhysique} of width of the order $n^{-1/2}$,
but in the limit of large $Z$, it gives a correction of lower order in $Z$).
The radius $R$ is determined by the perfect screening condition (\ref{A.6})
which gives $R^2=Z/(\pi n)$. The second moment is
\begin{equation} \label{A.7}
\int {\rm d}{\bf r}\:r^2\rho({\bf r}\vert Ze,{\bf 0})=-ne\pi R^4/2=
-Z^2\frac{e}{2\pi n}, \qquad Z\rightarrow\infty,
\end{equation} 
which is the $Z^2$ term of (\ref{2.1}), and this is the highest-order power of
$Z$ in the second moment.

The same argument extended to 3D gives
\begin{equation} \label{A.8}
\int {\rm d}{\bf r}\:r^2\rho({\bf r}\vert Ze,{\bf 0})=
-(3Z)^{5/3}\frac{e}{5(4\pi n)^{2/3}}, \qquad Z\rightarrow\infty.
\end{equation}
Therefore, in 3D, the second moment is \emph{not} a polynomial in $Z$, and no
exact formula valid for any $Z$ can be obtained by the present method.

\subsection{Guest charge in a two-component plasma}
A sum rule for the TCP can also be obtained from (\ref{2.12}). 
Now, we consider a mixture of three species, with charge $e_1=e$ and density
$n_1=n_+$, charge $e_2=-e$ and density $n_2=n_-$, charge $e_3=Ze$ and density 
$n_3$, respectively. 
There is no background ($\rho_b=0$). 
We choose $\alpha'=3$ in (\ref{2.12}). 
Finally, for dealing with one guest charge $Ze$ only, 
we set $n_3=0$, $n_+=n_-$ (neutrality); the system is stable against
collapse if $\beta e^2<2$ and $\beta Ze^2<2$. 
We call $n=n_++n_-$ the total density of the TCP. 
In (\ref{2.12}) appears the quantity 
\begin{equation} \label{2.14}
n_+h_{31}(r)+n_-h_{32}(r)=n({\bf r}\vert Ze,{\bf 0})-n, 
\end{equation}
which is the excess density around the guest charge $Ze$. 
Then, (\ref{2.12}) becomes a sum rule for the zeroth moment of 
this excess density: 
\begin{equation} \label{2.15}
\int{\rm d}{\bf r}\:[n({\bf r}\vert Ze,{\bf 0})-n]
=Z^2\frac{\beta e^2}{4-\beta e^2}. 
\end{equation}
This result is a generalization of the compressibility sum rule
\cite{Hill}
\begin{equation} \label{2.16}
\int{\rm d}{\bf r}\:[n({\bf r}\vert \pm e,{\bf 0})-n]
= \frac{\partial n}{\partial (\beta p)} -1
\end{equation}
with the use of the exact equation of state $\beta p = n(1-\beta e^2/4)$,
where $p$ is the pressure.

\subsection{Mixture without a background}
In the case $\rho_b=0$, another derivation of (\ref{2.12}) is possible
starting from the known equation of state \cite{SP}
\begin{equation}
\beta p=\sum_{\alpha}\left(1-\frac{\beta e_{\alpha}^2}{4}\right)n_{\alpha}.
\label{2.17}
\end{equation}
The $M$-component plasma may be described in the grand-canonical ensemble,
with $M$ chemical potentials $\mu_{\alpha}$ (actually \cite{LL}, the system
turns out to be neutral in the thermodynamic limit, and $M-1$ chemical
potentials would suffice for determining the state of the system; but here it
is more convenient to use $M$ chemical potentials). The pressure $p$ is given
by $\beta p=\lim(1/V)\ln\Xi$, where $V$ is the volume (here area) of the
system, $\Xi$ is the grand partition function, and $\lim$ is the thermodynamic 
limit. Taking the partial derivative of (\ref{2.17}) with respect to
$\beta\mu_{\alpha'}$ gives
\begin{equation}
n_{\alpha'}=\sum_{\alpha}\left(1-\frac{\beta e_{\alpha}^2}{4}\right)
\left(n_{\alpha}n_{\alpha'}\int{\rm d}{\bf r}\:h_{\alpha\alpha'}(r)\;
+n_{\alpha'}\delta_{\alpha,\alpha'}\right), \label{2.18}
\end{equation}
which is (\ref{2.12}) with $\rho_b=0$.

\renewcommand{\theequation}{3.\arabic{equation}}
\setcounter{equation}{0}

\section{Guest charge and potential fluctuations}
Putting a guest particle of charge $Ze$ at the origin ${\bf r}={\bf 0}$,
the original Hamiltonian $H_0$ of the infinite Coulomb system modifies
to $H = H_0+Ze \hat{\phi}({\bf 0})$, where $\hat{\phi}({\bf 0})$ is
the microscopic electric potential created at the origin by the Coulomb
system. The charge density around the guest charge, at point ${\bf r}$, is
thus expressible as 
\begin{equation} \label{3.1}
\rho({\bf r}\vert Ze,{\bf 0}) =
\frac{\langle \hat{\rho}({\bf r}) \exp\left[ -\beta Z e \hat{\phi}({\bf 0})
\right] \rangle}{\langle \exp\left[ -\beta Z e \hat{\phi}({\bf 0})
\right] \rangle} ,
\end{equation}
where $\langle\cdots\rangle$ denotes the thermal average over
the homogeneous system with the Hamiltonian $H_0$.

Let $\mu_{Ze}^{\rm ex}$ denotes the excess (i.e., over ideal)
chemical potential of the guest charge, i.e. the reversible work 
which has to be done to bring the guest particle of charge $Ze$ from 
infinity into the bulk interior of the considered Coulomb plasma.  
By the coupling parameter technique \cite{Hill}, this chemical potential 
can be represented in terms of the charge density (\ref{3.1}) 
as follows
\begin{equation} \label{3.2}
\mu_{Ze}^{\rm ex} = e \int_0^Z {\rm d}Z' \int {\rm d}{\bf r}~
v({\bf r}) \rho({\bf r}\vert Z'e,{\bf 0}) .
\end{equation}
With regard to the representation (\ref{3.1}),
$\mu_{Ze}^{\rm ex}$ can be expressed as
\begin{equation} \label{3.3}
-\beta \mu_{Ze}^{\rm ex} = \int_0^{-\beta Ze} {\rm d}x
\frac{\langle \hat{\phi} \exp(x\hat{\phi})\rangle}{\langle
\exp(x\hat{\phi}) \rangle} .
\end{equation}
Here, since the thermal averages are point-independent,
we use the notation $\hat{\phi} \equiv \hat{\phi}({\bf 0})$.

Let us recall some basic information about the cumulant expansion.
Let $\hat{\phi}$ be a random variable with the probability distribution
$P(\hat{\phi})$.
The cumulant expansion is defined by
\begin{equation} \label{3.4}
\langle \exp(x\hat{\phi}) \rangle = \exp\left(\sum_{l=1}^{\infty} 
\frac{x^l}{l!} \langle \hat{\phi}^l \rangle_c\right) ,
\end{equation}
where $x$ is any complex number and $\langle\hat{\phi}^l \rangle_c$ 
are the cumulants. 
They are combinations of the standard moments $\langle\hat{\phi}^l \rangle$.
Differentiating the equality (\ref{3.4}) with respect to $x$ gives
\begin{equation} \label{3.5}
\frac{{\rm d}}{{\rm d}x}
\sum_{l=0}^{\infty} \frac{x^l}{l!} \langle \hat{\phi}^l \rangle 
=\exp\left( \sum_{l=1}^{\infty} 
\frac{x^l}{l!} \langle \hat{\phi}^l \rangle_c \right)
\frac{{\rm d}}{{\rm d}x}\sum_{l=1}^{\infty}
\frac{x^l}{l!} \langle \hat{\phi}^l \rangle_c .
\end{equation}
Equating the coefficients of the same power of $x$ in both sides of
(\ref{3.5}) gives the recursion formula
\begin{equation} \label{3.6}
\langle\hat{\phi}^l \rangle_c = \langle\hat{\phi}^l \rangle
- \sum_{k=1}^{l-1} {l-1\choose k-1} \langle\hat{\phi}^k \rangle_c
\langle\hat{\phi}^{l-k} \rangle .  
\end{equation}
The first cumulants read
\begin{eqnarray}
\langle\hat{\phi} \rangle_c & = & \langle\hat{\phi} \rangle , \nonumber \\
\langle\hat{\phi}^2 \rangle_c & = & \langle\hat{\phi}^2 \rangle  
- \langle\hat{\phi}\rangle^2 , \label{3.7} \\
\langle\hat{\phi}^3 \rangle_c & = & \langle\hat{\phi}^3 \rangle  
- 3 \langle\hat{\phi}^2\rangle \langle\hat{\phi}\rangle 
+ 2\langle\hat{\phi}\rangle^3, 
\nonumber
\end{eqnarray}
etc.
In the theory of fluids, the cumulants of type (\ref{3.7})
are referred to as truncations, and therefore we shall use the notation
$\langle\hat{\phi}^l\rangle_c\equiv \langle\hat{\phi}^l\rangle^{\rm T}$.  

Since it holds
\begin{equation} \label{3.8}
\frac{\langle \hat{\phi} \exp(x\hat{\phi})\rangle}{\langle\exp(x\hat{\phi}) 
\rangle} = \frac{{\rm d}}{{\rm d}x} \ln \langle\exp(x\hat{\phi})\rangle ,
\end{equation}
the excess chemical potential (\ref{3.3}) is expressible as
\begin{equation} \label{3.9}
- \beta \mu_{Ze}^{\rm ex} = \ln \langle\exp(-\beta Ze\hat{\phi})\rangle . 
\end{equation}
Based on the recapitulation in the above paragraph, $\mu_{Ze}^{\rm ex}$
is expressible either in the form of a cumulant expansion
\begin{equation} \label{3.10}
-\beta \mu_{Ze}^{\rm ex} = \sum_{l=1}^{\infty} 
\frac{(-\beta Ze)^l}{l!} \langle \hat{\phi}^l \rangle^{\rm T} ,
\end{equation}
or in the form of the standard moment expansion
\begin{equation} \label{3.11}
\exp\left( -\beta \mu_{Ze}^{\rm ex}\right) = 
\langle\exp(-\beta Ze\hat{\phi})\rangle \equiv 
1 + \sum_{l=1}^{\infty} \frac{(-\beta Ze)^l}{l!} 
\langle \hat{\phi}^l \rangle .
\end{equation}
It stands to reason that the expansions (\ref{3.10}) and (\ref{3.11}) 
are valid provided all moments exist.
We conclude that the knowledge of the excess chemical potential of
the guest particle with an arbitrary charge provides the exact
information about all moments of the electrostatic potential
at a point of the infinite homogeneous Coulomb system.

Going to the infinite system via the thermodynamic limit of a finite system 
with a disc geometry \cite{Alastuey84}, the fluctuations of the potential
at any point become infinite due to the presence of dipoles near the boundary.
Here, the potential moments are defined directly for an infinite space,
without the presence of a boundary.
This corresponds to going to the infinite system via the thermodynamic limit 
of a finite system, e.g., with periodic boundary conditions, formulated on 
the surface of a sphere and so on. 
We thus expect that the average potential at a point is equal to zero and 
all its moments are finite. 
 
Since in 2D the potential (\ref{1.2}) is dimensionless,
$\hat{\phi}$ has the dimension of the elementary charge $e$.
It is therefore useful to introduce the dimensionless microscopic 
quantity $\psi = \hat{\phi}/e$ with the probability 
distribution $P(\psi)$.
Setting in (\ref{3.11}) $\beta Ze^2 = {\rm i}k$, one gets
\begin{equation} \label{3.12}
\exp\left( -\beta \mu_{Ze}^{\rm ex}\right) \big\vert_{\beta Ze^2={\rm i}k}
= \langle\exp(-{\rm i}k\psi)\rangle =
\int_{-\infty}^{\infty} {\rm d}\psi~{\rm e}^{-{\rm i}k\psi}
P(\psi) \equiv \tilde{P}(k) ,
\end{equation}
where $\tilde{P}(k)$ is the Fourier component of the $\psi$-distribution.
The original probability distribution $P(\psi)$ can be obtained by 
the Fourier inversion of this relation
\begin{equation} \label{3.13}
P(\psi) = \int_{-\infty}^{\infty}\frac{{\rm d}k}{2\pi}{\rm e}^{{\rm i}k\psi}
\exp\left( -\beta \mu_{Ze}^{\rm ex}\right) \big\vert_{\beta Ze^2={\rm i}k} .
\end{equation}

All that has been said in this section is valid also for $v$ being
the pure Coulomb potential plus any type of short-distance 
regularization. 

\renewcommand{\theequation}{4.\arabic{equation}}
\setcounter{equation}{0}

\section{High-temperature limit}
The high-temperature (weak-coupling) limit of Coulomb systems
is described rigorously by the Debye-H\"uckel theory \cite{Debye,Kennedy}.
In 2D, the two-body Ursell functions $U$ of charged species 
are given by \cite{Jancovici04}
\begin{equation} \label{4.1}
U_{\alpha\alpha'}({\bf r},{\bf r}') \equiv
n_{\alpha\alpha'}^{(2)}({\bf r},{\bf r}') - n_{\alpha} n_{\alpha'}
= - e_{\alpha} n_{\alpha} e_{\alpha'} n_{\alpha'} 
\beta K_0(\kappa\vert {\bf r}-{\bf r}'\vert) ,
\end{equation}
where $K_0$ is a modified Bessel function \cite{Gradshteyn} and
$\kappa = (2\pi\beta\sum_{\alpha}e_{\alpha}^2 n_{\alpha})^{1/2}$
is the inverse Debye length.

The potential-potential correlation function can be calculated directly
from the definition
\begin{eqnarray}
\langle \hat{\phi}({\bf 0}) \hat{\phi}({\bf r}) \rangle^{\rm T} & = &
\int {\rm d}{\bf r}_1~v({\bf r}-{\bf r}_1) \int {\rm d}{\bf r}_2~v({\bf r}_2)
\langle \hat{\rho}({\bf r}_1) \hat{\rho}({\bf r}_2) \rangle^{\rm T} 
\nonumber \\ & = & \int {\rm d}{\bf r}_1~v({\bf r}-{\bf r}_1) 
\int {\rm d}{\bf r}_2~v({\bf r}_1-{\bf r}_2)
\langle \hat{\rho}({\bf 0}) \hat{\rho}({\bf r}_2) \rangle^{\rm T} . \label{4.2}
\end{eqnarray}
Using for the Coulomb potential the expansion in polar coordinates
\begin{equation} \label{4.3}
v({\bf r}_1-{\bf r}_2) = - \ln\vert {\bf r}_1-{\bf r}_2\vert
= - \ln r_> + \sum_{l=1}^{\infty} \frac{1}{l}
\left( \frac{r_<}{r_>} \right)^l \cos l(\theta_1-\theta_2) 
\end{equation}
with $r_< = \min\{ r_1,r_2\}$ and $r_> = \max\{ r_1,r_2\}$, and
taking into account the screening sum rule \cite{Martin}
\begin{equation} \label{4.4}
\int {\rm d}{\bf r}_2~\langle \hat{\rho}({\bf 0}) \hat{\rho}({\bf r}_2) 
\rangle^{\rm T} = 0 ,
\end{equation}
the second integral on the rhs of (\ref{4.2}) can be expressed as
\begin{equation} \label{4.5}
\int {\rm d}{\bf r}_2~v({\bf r}_1-{\bf r}_2)
\langle \hat{\rho}({\bf 0}) \hat{\rho}({\bf r}_2) \rangle^{\rm T} =
- \int_{r_1}^{\infty} {\rm d}r_2~2\pi r_2 \ln\left( \frac{r_2}{r_1} \right)
\langle \hat{\rho}({\bf 0}) \hat{\rho}({\bf r}_2) \rangle^{\rm T} . 
\end{equation}
Considering the charge correlation function
\begin{eqnarray}
\langle \hat{\rho}({\bf 0}) \hat{\rho}({\bf r}_2) \rangle^{\rm T} & = &
\sum_{\alpha,\alpha'} e_{\alpha} e_{\alpha'} \left[
U_{\alpha\alpha'}^{(2)}({\bf r}_2) + n_{\alpha} \delta_{\alpha\alpha'}
\delta({\bf r}_2) \right] \nonumber \\ & = &
- \frac{\kappa^4}{(2\pi)^2\beta} K_0(\kappa r_2)
+ \frac{\kappa^2}{2\pi\beta} \delta({\bf r}_2) \label{4.6}
\end{eqnarray}
in equation (\ref{4.5}) implies, after an integration by parts,
\begin{equation} \label{4.7}
\int {\rm d}{\bf r}_2~v({\bf r}_1-{\bf r}_2)
\langle \hat{\rho}({\bf 0}) \hat{\rho}({\bf r}_2) \rangle^{\rm T} =
\frac{\kappa^2}{2\pi\beta} K_0(\kappa r_1) .
\end{equation}
Inserting this relation into (\ref{4.2}) and applying once more
the expansion (\ref{4.3}) results into
\begin{equation} \label{4.8}
\beta \langle \hat{\phi}({\bf 0}) \hat{\phi}({\bf r}) \rangle^{\rm T} 
= - \ln r - K_0(\kappa r) .
\end{equation}
This procedure will be repeated, without going into details,
also in the cases treated in the next sections.

The result (\ref{4.8}) has the correct large-distance asymptotic
\cite{LeboMartin}
\begin{equation} \label{4.9}
\beta \langle \hat{\phi}({\bf 0}) \hat{\phi}({\bf r}) \rangle^{\rm T} 
\mathop{\sim}_{r\to\infty} - \ln r .
\end{equation}
In the zero-distance limit $r\to 0$, using the expansion
\begin{equation} \label{4.10}
K_0(x) = - C - \ln (x/2) + O(x^2\ln x)
\end{equation}
with $C$ being the Euler number, the one-point second-moment 
fluctuation formula for the potential reads
\begin{equation} \label{4.11}
\beta \langle \hat{\phi}^2 \rangle^{\rm T} = C + \ln(\kappa/2) .
\end{equation}

One can obtain the last result in an alternative way by considering
the charge density induced around the guest charge \cite{Samaj07}
\begin{equation} \label{4.12}
\rho({\bf r}\vert Ze,{\bf 0}) = - Z e \frac{\kappa^2}{2\pi} K_0(\kappa r) .
\end{equation}
Then, according to (\ref{3.2}),
\begin{eqnarray}
-\beta \mu_{Ze}^{\rm ex} & = & -\beta e^2 \int_0^Z {\rm d}Z'~Z' 
\frac{\kappa^2}{2\pi} \int_0^{\infty} {\rm d}r~2\pi r \ln r K_0(\kappa r) 
\nonumber \\ & = &
\frac{\beta (Ze)^2}{2} \left[ C + \ln(\kappa/2) \right] . \label{4.13}
\end{eqnarray}
With regard to the cumulant expansion (\ref{3.10}), we recover
the previous result (\ref{4.11}).

From (\ref{3.10}) and (\ref{4.13}), all the higher-order truncated moments 
$\langle \hat{\phi}^l \rangle^{\rm T}$ with $l\ge 3$ vanish in 
the Debye-H\"uckel limit; this indicates a Gaussian
distribution for the one-point potential in this limit.
We shall return to this problem and present all truncated potential moments,
for the TCP, in a high-temperature limit going beyond the the Debye-H\"uckel
limit, in Sect. 6.

\renewcommand{\theequation}{5.\arabic{equation}}
\setcounter{equation}{0}

\section{One-component plasma at $\beta e^2 = 2$}
The 2D OCP is exactly solvable in terms of free-fermions when the 
dimensionless coupling constant $\beta e^2$ has the special value 2 
\cite{Alastuey81,Jancovici81}.
In the thermodynamic limit, the two-body Ursell function of mobile
particles at distance $r$ is 
\begin{equation} \label{5.1}
U(r) = - n^2 \exp\left( -\pi n r^2 \right) ,
\end{equation}
where $n$ is the particle density.
All many-body Ursell functions are known at the free-fermion point, too.

The potential-potential correlation function can be calculated
in close analogy with the previous steps outlined between
Eqs. (\ref{4.2})-(\ref{4.8}).
Substituting the charge correlation function
\begin{equation} \label{5.2}
\langle \hat{\rho}({\bf 0}) \hat{\rho}({\bf r}_2) \rangle^{\rm T} =
- e^2 n^2 \exp(-\pi n r_2^2) + n \delta({\bf r}_2)
\end{equation}
into the relation (\ref{4.5}) and using an integration by parts,
one gets
\begin{equation} \label{5.3}
\int {\rm d}{\bf r}_2~v({\bf r}_1-{\bf r}_2)
\langle \hat{\rho}({\bf 0}) \hat{\rho}({\bf r}_2) \rangle^{\rm T} =
\frac{e^2 n}{2} \Gamma(0,\pi n r_1^2) ,
\end{equation}
where
\begin{equation} \label{5.4}
\Gamma(x,t) = \int_t^{\infty} {\rm d}s~s^{x-1} {\rm e}^{-s}
\end{equation}
is the incomplete Gamma function.
From (\ref{4.2}), one thus obtains
\begin{equation} \label{5.5}
\beta \langle \hat{\phi}({\bf 0}) \hat{\phi}({\bf r}) \rangle^{\rm T}
= - \ln r + \frac{1}{2} \left[ {\rm e}^{-\pi n r^2} -
(1+\pi n r^2) \Gamma(0,\pi n r^2) \right] .
\end{equation} 
This result has the correct large-distance asymptotic (\ref{4.9}).
In the zero-distance limit, it yields
\begin{equation} \label{5.6}
\beta \langle \hat{\phi}^2 \rangle^{\rm T} =
\frac{1}{2} \left[ 1 + C + \ln(\pi n) \right] .
\end{equation}
Note that the large-distance behavior (\ref{4.9}) is universal, while the
zero-distance limit (\ref{4.11}) or (\ref{5.6}) depends on the coupling
constant $\beta e^2$. 

All potential moments are available for the present system
due to the knowledge of the induced charge density around
the guest charge \cite{Jancovici84,Samaj07}:
\begin{equation} \label{5.7}
\rho({\bf r}\vert Ze,{\bf 0}) = - e n
\frac{\Gamma(Z,\pi n r^2)}{\Gamma(Z)} , \qquad Z\ge 0 .
\end{equation} 
By using the relation (\ref{3.2}), one obtains after some algebra
\cite{Jancovici84}
\begin{equation} \label{5.8}
- \beta \mu_{Ze}^{\rm ex} = \frac{Z^2}{2} \left[ 1 + \ln(\pi n) \right]
- \int_0^Z {\rm d}Z'~Z' \psi(1+Z') ,
\end{equation}
where $\psi$ is the psi-function defined by
\begin{equation} \label{5.9}
\psi(x) = \frac{{\rm d}}{{\rm d}x} \ln \Gamma(x) .
\end{equation}
Its Taylor expansion around $x=1$ reads \cite{Gradshteyn}
\begin{equation} \label{5.10}
\psi(1+x) = - C + \sum_{l=2}^{\infty} (-1)^l \zeta(l) x^{l-1} ,
\end{equation}
where
\begin{equation} \label{5.11}
\zeta(l) = \sum_{k=1}^{\infty} \frac{1}{k^l}
\end{equation}
is the Riemann zeta function.
Considering the expansion (\ref{5.10}) in (\ref{5.8}) gives
\begin{equation} \label{5.12}
-\beta \mu_{Ze}^{\rm ex} = \frac{Z^2}{2} \left[ 1 + C + \ln(\pi n) \right]
+ \sum_{l=3}^{\infty} \frac{(-1)^l Z^l}{l} \zeta(l-1) .
\end{equation}
The comparison of this expansion with the cumulant expansion (\ref{3.10})
implies
\begin{eqnarray}
\langle \hat{\phi}^2 \rangle^{\rm T} & = & \frac{e^2}{4}
\left[ 1 + C + \ln(\pi n) \right] , \label{5.13} \\
\langle \hat{\phi}^l \rangle^{\rm T} & = & \frac{e^l}{2^l}
(l-1)! \zeta(l-1) , \qquad l\ge 3 . \label{5.14}
\end{eqnarray}
Note that the second-moment formula (\ref{5.13}) is identical to the previous
one (\ref{5.6}) derived by the direct calculation from the definition.

\renewcommand{\theequation}{6.\arabic{equation}}
\setcounter{equation}{0}

\section{Two-component plasma}

\subsection{Collapse point $\beta e^2 = 2$}
The 2D TCP of $\pm e$ charges is mappable for the special value
of the coupling constant $\beta e^2 = 2$ onto the Thirring model
at the free-fermion point \cite{Cornu87,Cornu89}.
Although this coupling corresponds to the collapse threshold
for the pointlike particles, and therefore for a fixed fugacity $z$
the particle density is infinite, the Ursell functions are well defined.
Their two-body forms read
\begin{equation} \label{6.1}
U_{\pm,\pm}(r) = - \left( \frac{m^2}{2\pi} \right)^2
K_0^2(m r) , \qquad
U_{\pm,\mp}(r) = \left( \frac{m^2}{2\pi} \right)^2
K_1^2(m r) ,
\end{equation}
where $m=2\pi z$.
All many-body Ursell functions are also known.

Substituting the charge correlation function
\begin{equation} \label{6.2}
\langle \hat{\rho}({\bf 0}) \hat{\rho}({\bf r}_2) \rangle^{\rm T} =
- 2 e^2 \left( \frac{m^2}{2\pi} \right)^2 
\left[ K_0^2(m r_2) + K_1^2(m r_2) \right]
\end{equation}
into the relation (\ref{4.5}) and integrating by parts leads to
\begin{equation} \label{6.3}
\int {\rm d}{\bf r}_2~v({\bf r}_1-{\bf r}_2)
\langle \hat{\rho}({\bf 0}) \hat{\rho}({\bf r}_2) \rangle^{\rm T} =
e^2 \frac{m^2}{2\pi} K_0^2(m r_1) .
\end{equation}
From (\ref{4.2}), one finds that
\begin{equation} \label{6.4}
\beta \langle \hat{\phi}({\bf 0}) \hat{\phi}({\bf r}) \rangle^{\rm T}
= - \ln r + \frac{(m r)^2}{2} \left[ 2 K_1^2(m r) - K_0^2(m r)
- K_0(m r) K_2(m r) \right] .
\end{equation} 
This result has the correct large-distance asymptotic (\ref{4.9}).
In the zero-distance limit, it gives
\begin{equation} \label{6.5}
\beta \langle \hat{\phi}^2 \rangle^{\rm T} = 1 + C + \ln(\pi z) .
\end{equation}

\subsection{Stability region $0\le \beta e^2 < 2$}
The system of pointlike $\pm e$ charged particles is stable against
the collapse of positive-negative pairs of charges provided that
the corresponding Boltzmann weight 
$\exp[\beta e^2 v({\bf r})] = r^{-\beta e^2}$ can be integrated
at short 2D distances, i.e. when $\beta e^2<2$.
The equilibrium statistical mechanics of the neutral TCP is usually
studied in the grand canonical ensemble, characterized by the particle
fugacities $z_+ = z_- = z$. The full thermodynamics of this system is 
known \cite{Samaj00,Samaj03}.

In the stability range of $\beta e^2<2$, the grand partition function
$\Xi(z)$ of the 2D TCP can be turned via the Hubbard-Stratonovich
transformation (see, e.g., Ref. \cite{Minnhagen}) into
\begin{equation} \label{6.6}
\Xi(z) = \frac{\int {\cal D}\varphi \exp[-S(z)]}{\int {\cal D}\varphi 
\exp[-S(0)]} ,
\end{equation}
where
\begin{equation} \label{6.7}
S(z) = \int {\rm d}{\bf r} \left[ \frac{1}{16\pi} (\nabla\varphi)^2
- 2 z \cos( b\varphi ) \right] 
\end{equation}
is the Euclidean action of the $(1+1)$-dimensional sine-Gordon model. 
Here, $\varphi({\bf r})$ is a real scalar field and $\int {\cal D}\varphi$
denotes the functional integration over this field. 
The sine-Gordon coupling constant $b$ depends on the Coulomb coupling 
constant via
\begin{equation} \label{6.8}
b = \sqrt{\frac{\beta e^2}{4}} .
\end{equation}
The fugacity $z$ is renormalized by the diverging self-energy term
$\exp[\beta v({\bf 0})/2]$ which disappears from statistical relations
under the conformal short-distance normalization of the exponential fields
\cite{Samaj00,Samaj03}
\begin{equation} \label{6.9}
\langle {\rm e}^{{\rm i}b\varphi({\bf r})} {\rm e}^{-{\rm i}b\varphi({\bf r}')}
\rangle_{\rm sG} \sim \vert {\bf r}-{\bf r}'\vert^{-4 b^2}
\qquad \mbox{as $\vert {\bf r}-{\bf r}' \vert \to 0$,}
\end{equation}
where $\langle \cdots \rangle_{\rm sG}$ denotes the average with
the sine-Gordon action (\ref{6.7}).
The species densities are expressible in the sine-Gordon format as follows
\begin{equation} \label{6.10}
n_{\pm} = z \langle {\rm e}^{\pm{\rm i}b\varphi} \rangle_{\rm sG} .
\end{equation}
The charge neutrality of the system $n_+ = n_- = n/2$ is ensured
by the obvious symmetry relation
$\langle {\rm e}^{{\rm i}b\varphi}\rangle_{\rm sG} =
\langle {\rm e}^{-{\rm i}b\varphi}\rangle_{\rm sG}$.

The excess chemical potential of the particle species forming the plasma
is given by
\begin{equation} \label{6.11}
\exp(-\beta \mu_{\pm e}^{\rm ex}) = \frac{n_{\pm}}{z}
= \langle {\rm e}^{\pm{\rm i}b\varphi} \rangle_{\rm sG} .
\end{equation}
It was shown in Ref. \cite{Samaj05} that the excess chemical potential
of a guest charge $Ze$ immersed in the plasma is expressible in 
the sine-Gordon format as follows
\begin{equation} \label{6.12}
\exp(-\beta \mu_{Z e}^{\rm ex}) = 
\langle {\rm e}^{{\rm i}Z b\varphi} \rangle_{\rm sG} .
\end{equation}
When $Z=\pm 1$, one recovers the previous result (\ref{6.11}) valid
for the plasma constituents.
Due to the symmetry relation
$\langle {\rm e}^{{\rm i}a\varphi} \rangle_{\rm sG} =
\langle {\rm e}^{-{\rm i}a\varphi} \rangle_{\rm sG}$ valid for any
real-valued $a$, it holds that
$\mu_{Z e}^{\rm ex} =  \mu_{-Z e}^{\rm ex}$.  

The (1+1)-dimensional sine-Gordon model is an integrable
field theory \cite{Zamolodchikov79}.
Due to a recent progress in the method of the Thermodynamic Bethe ansatz,
a general formula for the expectation value of the exponential field
$\langle {\rm e}^{{\rm i}a\phi}\rangle$ was derived by 
Lukyanov and Zamolodchikov \cite{Lukyanov}.
In the notation of equation (\ref{6.12}), $a = Z b$, their formula reads
\begin{equation} \label{6.13}
\langle {\rm e}^{{\rm i}Z b \varphi} \rangle_{\rm sG} =
\left[ \frac{\pi z \Gamma(1-b^2)}{\Gamma(b^2)} \right]^{(Z b)^2/(1-b^2)} 
\exp\left[ I_b(Z) \right] 
\end{equation}
with
\begin{equation} \label{6.14}
I_b(Z) = \int_0^{\infty} \frac{{\rm d}t}{t} \left[
\frac{\sinh^2(2 Z b^2 t)}{2 \sinh(b^2 t) \sinh(t)
\cosh[(1-b^2)t]} - 2 Z^2 b^2 {\rm e}^{-2 t} \right] .
\end{equation} 
The interaction Boltzmann factor of the guest charge $Ze$ 
with an opposite plasma counterion at distance $r$, 
$r^{-\beta e^2\vert Z\vert}$, is integrable at small 2D
distances $r$ if $\beta \vert Z\vert e^2<2$, i.e. $\vert Z\vert < 1/(2 b^2)$;
this is indeed the condition for the integral (\ref{6.14}) to be finite, so 
that the couple of Eqs. (\ref{6.13}) and (\ref{6.14}) passes the collapse test.
Finally, using eqs. (\ref{6.13}) and (\ref{6.14}) in (\ref{6.12}),
one arrives at
\begin{equation} \label{6.15}
- \beta \mu_{Ze}^{\rm ex} = Z^2 \frac{b^2}{1-b^2}
\ln \left[ \frac{\pi z \Gamma(1-b^2)}{\Gamma(b^2)} \right] + I_b(Z) .
\end{equation}
We have to keep in mind that $b^2=\beta e^2/4$.

Comparing the cumulant expansion (\ref{3.10}) with the result
(\ref{6.15}), in which the integral $I_b(Z)$ (\ref{6.14}) is expanded
in powers of $Z$, one gets the explicit forms of the potential moments:
\begin{eqnarray}
\langle \hat{\phi}^2 \rangle^{\rm T} & = & \frac{e^2}{8 b^2 (1-b^2)} 
\ln \left[ \frac{\pi z \Gamma(1-b^2)}{\Gamma(b^2)} \right] \nonumber \\
& & + \frac{e^2}{4} \int_0^{\infty} \frac{{\rm d}t}{t}
\left[ \frac{t^2}{\sinh(b^2 t) \sinh(t) \cosh[(1-b^2)t]}
- \frac{1}{b^2} {\rm e}^{-2 t} \right] , \label{6.16} \\
\langle \hat{\phi}^{2l} \rangle^{\rm T} & = & \frac{e^{2l}}{4}
\int_0^{\infty} {\rm d}t
\frac{t^{2l-1}}{\sinh(b^2 t) \sinh(t) \cosh[(1-b^2)t]} ,
\quad l=2,3,\ldots . \phantom{aa} \label{6.17} 
\end{eqnarray}  
The odd potential moments vanish for the symmetric TCP.

In the high-temperature limit $\beta e^2\to 0$ $(b^2\to 0)$, 
(\ref{6.15}) taken with $z\sim n/2$ reduces to the previous one (\ref{4.13});
one retrieves the second moment (\ref{4.11}) and that all higher moments
vanish, as it should be. From  (\ref{6.17}), in the limit $b^2\rightarrow 0$,
one finds
\begin{equation} \label{6.18}
\beta \langle \hat{\phi}^{2l} \rangle^{\rm T} = e^{2(l-1)} 8
\frac{4^l-2}{4^{2l}} (2l-2)! \zeta(2l-1) , \qquad l=2,3,\ldots .
\end{equation}
These expressions go beyond the Debye-H\"uckel limit of (\ref{6.15}).

At the collapse point $\beta e^2=2$ $(b^2=1/2)$, the second-moment formula
(\ref{6.16}) reproduces the previous result (\ref{6.5}) and the higher-order
moments (\ref{6.17}) take forms
\begin{equation} \label{6.19}
\langle \hat{\phi}^{2l} \rangle^{\rm T} = e^{2l} 
\frac{2}{4^l} (2l-1)! \zeta(2l-1) , \qquad l=2,3,\ldots .
\end{equation}

All potential moments are finite also in the collapse region, up to
the Kosterlitz-Thouless critical point $\beta e^2=4$ $(b^2=1)$.
We conjecture that, in the case of the hard-core regularization of the
Coulomb potential, the obtained result correspond to the limit of
a vanishing hard core.

We end up this section by a comment about the possibility of a relationship
between the electrostatic potential $\hat{\phi}$ and the sine-Gordon field
variable $\varphi$.
This relationship was suggested in many articles, see, e.g., 
Ref. \cite{Dean}.
The comparison of Eqs. (\ref{3.11}) and (\ref{6.12}) implies
\begin{equation} \label{6.20}
\langle \varphi^{2l} \rangle_{\rm sG} = (-1)^l (4\beta)^l
\langle \hat{\phi}^{2l} \rangle . 
\end{equation}
This means that, in view of one-point fluctuations, the fields 
$\hat{\phi}$ and $\varphi$ differ from one another only by 
an irrelevant scaling factor.
On the other hand, the large-distance asymptotic of 
the potential-potential correlations (\ref{4.9}) is fundamentally
different from the one of 
$\langle \varphi({\bf 0}) \varphi({\bf r}) \rangle^{\rm T}$
The latter two-point correlation function has, 
like in every massive field theory, 
a short-range exponential decay as $r\to\infty$.
We conclude that the electrostatic-potential interpretation of the
sine-Gordon field is not correct.

\renewcommand{\theequation}{7.\arabic{equation}}
\setcounter{equation}{0}

\section{Conclusion}
The general study of a mixture of $M$ species of mobile particles, which may
be embedded in a uniform background, is simpler in two dimensions; the BGY
hierarchy suffices for deriving the general sum rule (\ref{2.12}) relating the
second moments and the zeroth moments of the two-body correlations. Further
work should be possible about this mixture.

\section*{Acknowledgments}
B. Jancovici has benefited of a stimulating conversation with
L. Suttorp.\linebreak 
L. \v{S}amaj is grateful to LPT for its very kind invitation;
the support by grant VEGA 2/6071/27 is acknowledged.


\newpage

\begin{thebibliography}{99}

\bibitem{Samaj07} \v{S}amaj, L.:
J. Stat. Phys. {\bf 128}, 1415 (2007)

\bibitem{Suttorp} Suttorp, L.G., van Wonderen, A.J.:
Physica A {\bf 145}, 533 (1987)

\bibitem{Martin} Martin, Ph.A.:
Rev. Mod. Phys. {\bf 60}, 1075 (1988)

\bibitem{Gruber} Gruber, Ch., Lebowitz, J.L., Martin, Ph.A.:
J. Chem. Phys. {\bf 75}, 944 (1981)

\bibitem{Hauge} Hauge, E.H., Hemmer, P.C.:
Phys. Norv. {\bf 5}, 109 (1971), and references quoted there

\bibitem{JancoviciPhysique} Jancovici, B.:
J. Physique-Lettres {\bf 42}, L-223 (1981)

\bibitem{Hill} See, e.g., Hill, T.L.:
Statistical Mechanics. McGraw-Hill (1956)

\bibitem{SP} Salzberg, A.M., Prager, S.:
J. Chem. Phys. {\bf 38}, 2587 (1963)

\bibitem{LL} Lieb, E.H., Lebowitz, J.L.,:
Adv. in Math. {\bf 9}, 316 (1972)

\bibitem{Alastuey84} Alastuey, A., Jancovici, B.:
J. Stat. Phys. {\bf 34}, 557 (1984)

\bibitem{Debye} Debye, P., H\"uckel, E.:
Phys. Z. {\bf 24}, 185 (1923)

\bibitem{Kennedy} Kennedy, T.: 
Comm. Math. Phys. {\bf 92}, 269 (1983) 

\bibitem{Jancovici04} Jancovici, B., \v{S}amaj, L.:
J. Stat. Phys. {\bf 114}, 1211 (2004)

\bibitem{Gradshteyn} Gradshteyn, I.S., Ryzhik, I.M.:
Table of Integrals, Series and Products, 5th ed.
Academic Press, London, (1994)

\bibitem{LeboMartin} Lebowitz, J.L., Martin, Ph.A.:
J. Stat. Phys. {\bf 34}, 287 (1984)

\bibitem{Alastuey81} Alastuey, A., Jancovici, B.:
J. Phys. (Paris) {\bf 42}, 1 (1981)

\bibitem{Jancovici81} Jancovici, B.:
Phys. Rev. Lett. {\bf 46}, 386 (1981)

\bibitem{Jancovici84} Jancovici, B.:
Mol. Phys. {\bf 52}, 1251 (1984)  

\bibitem{Cornu87} Cornu, F., Jancovici, B.:
J. Stat. Phys. {\bf 49}, 33 (1987)

\bibitem{Cornu89} Cornu, F., Jancovici, B.:
J. Chem. Phys. {\bf 90}, 2444 (1989)

\bibitem{Samaj00} \v{S}amaj, L., Trav\v{e}nec. I.:
J. Stat. Phys. {\bf 101}, 713 (2000)

\bibitem{Samaj03} \v{S}amaj, L.:
J. Phys. A: Math. Gen. {\bf 36}, 5913 (2003)

\bibitem{Minnhagen} Minnhagen, P.:
Rev. Mod. Phys. {\bf 59}, 1001 (1987)

\bibitem{Samaj05} \v{S}amaj, L.:
J. Stat. Phys. {\bf 120}, 125 (2005)

\bibitem{Zamolodchikov79} Zamolodchikov, A., Zamolodchikov, Al.:
Ann. Phys. (N.Y.) {\bf 120}, 253 (1979)

\bibitem{Lukyanov} Lukyanov, S., Zamolodchikov, Al.:
Nucl. Phys. B {\bf 493}, 571 (1997)

\bibitem{Dean} Dean, D.S., Horgan, R.R.: 
Phys. Rev. E {\bf 68}, 061106 (2003)

\end{thebibliography}
\end{document}